\documentclass[]{mn2e}
\usepackage{natbib}
\usepackage{epsfig}
\usepackage{amssymb}
\usepackage{amsmath}
\usepackage{graphicx}
\usepackage{tabularx}
\usepackage{rotfloat}
\usepackage{bm}
\usepackage{multirow}
\newcommand{\spcfs}{$\xi_s(\sigma,\pi)$}

\newcommand{\cfr}{$\xi_r$}

\begin{document}

\title[The growth rate in WiggleZ]{The WiggleZ Dark Energy Survey:
  measuring the cosmic growth rate with the two-point galaxy correlation
  function}

\author[Contreras et al.]{\parbox[t]{\textwidth}{Carlos Contreras$^1$,
    Chris Blake$^1$, Gregory B.\ Poole$^1$, Felipe Marin$^1$, Sarah Brough$^2$, 
    Matthew Colless$^2$, Warrick Couch$^1$, Scott Croom$^3$, Darren Croton$^1$, Tamara M.\ Davis$^4$,
    Michael J.\ Drinkwater$^4$, Karl Forster$^5$, David Gilbank$^6$, Mike Gladders$^7$, Karl Glazebrook$^1$,
    Ben Jelliffe$^3$, Russell J. Jurek$^8$,
    I-hui Li$^1$, Barry Madore$^9$, D. Christopher Martin$^5$, 
    Kevin Pimbblet$^{10}$, Michael Pracy$^{3}$, Rob Sharp$^{2,11}$, Emily Wisnioski$^1$, David Woods$^{12}$,
    Ted K.\ Wyder$^5$ and H.K.C. Yee$^{13}$} \\ \\ $^1$ Centre for
  Astrophysics \& Supercomputing, Swinburne University of Technology,
  P.O.\ Box 218, Hawthorn, VIC 3122, Australia \\ $^2$ Australian
  Astronomical Observatory, P.O. Box 296, Epping, NSW 1710, Australia
  \\ $^3$ Sydney Institute for Astronomy, School of Physics,
  University of Sydney, NSW 2006, Australia \\ $^4$ School of
  Mathematics and Physics, University of Queensland, Brisbane, QLD
  4072, Australia \\ $^5$ California Institute of Technology, MC
  278-17, 1200 East California Boulevard, Pasadena, CA 91125, United
  States \\ $^6$ South African Astronomical Observatory, PO Box 9, 
  Observatory, 7935, South Africa \\ $^7$ Department of Astronomy and 
  Astrophysics, University
  of Chicago, 5640 South Ellis Avenue, Chicago, IL 60637, United
  States \\ $^8$ Australia Telescope National Facility, CSIRO, Epping,
  NSW 1710, Australia \\ $^{9}$ Observatories of the Carnegie
  Institute of Washington, 813 Santa Barbara St., Pasadena, CA 91101,
  United States \\ $^{10}$ School of Physics, Monash University,
  Clayton, VIC 3800, Australia \\ $^{11}$ Research School of Astronomy
  \& Astrophysics, Australian National University, Weston Creek, ACT
  2611, Australia \\ $^{12}$ Department of Physics \& Astronomy,
  University of British Columbia, 6224 Agricultural Road, Vancouver,
  BC V6T 1Z1, Canada \\ $^{13}$ Department of Astronomy and
  Astrophysics, University of Toronto, 50 St.\ George Street, Toronto,
  ON M5S 3H4, Canada}

\maketitle

\begin{abstract}
The growth history of large-scale structure in the Universe is a
powerful probe of the cosmological model, including the nature of dark
energy.  We study the growth rate of cosmic structure to redshift $z =
0.9$ using more than $162{,}000$ galaxy redshifts from the WiggleZ
Dark Energy Survey.  We divide the data into four redshift slices with
effective redshifts $z = [0.2,0.4,0.6,0.76]$ and in each of the
samples measure and model the 2-point galaxy correlation function in
parallel and transverse directions to the line-of-sight.  After
simultaneously fitting for the galaxy bias factor we recover values
for the cosmic growth rate which are consistent with our assumed
$\Lambda$CDM input cosmological model, with an accuracy of around
$20\%$ in each redshift slice.  We investigate the sensitivity of our
results to the details of the assumed model and the range of physical
scales fitted, making close comparison with a set of N-body
simulations for calibration.  Our measurements are consistent with an
independent power-spectrum analysis of a similar dataset,
demonstrating that the results are not driven by systematic errors.
We determine the pairwise velocity dispersion of the sample in a
non-parametric manner, showing that it systematically increases with
decreasing redshift, and investigate the Alcock-Paczynski effects of
changing the assumed fiducial model on the results.  Our techniques
should prove useful for current and future galaxy surveys mapping the
growth rate of structure using the 2-dimensional correlation function.
\end{abstract}

\section{Introduction}
\label{sec:intro}

The clustering pattern of galaxies, which is driven by the competing
forces of gravitational attraction and universal cosmic expansion, is
one of the most important probes of the cosmological model. In
particular, measurements of the growth rate of cosmic structure,
inferred from the patterns imprinted in the galaxy clustering
distribution, can potentially distinguish between the increasing
number of alternative theories of gravity which currently compete with
General Relativity to explain the observed acceleration of the
universal expansion without the need for an exotic component like Dark
Energy \citep[see][for a review]{Tsu10}.  In general the cosmic growth
rate will evolve differently in these models, even if they have the
same expansion history, allowing constraints to be placed on different
theories of gravity \citep{LJ03,LC07,G08,W08}.

It has long been known in standard gravity that the growth rate $f =
d\ln{D}/d\ln{a}$ (in terms of the linear growth factor $D$ and cosmic
scale factor $a$) is well-approximated by $f(z) \approx
\Omega_m(z)^{0.6}$, where $\Omega_m(z)$ is the matter density relative
to critical density at redshift $z$ \citep{Pb80}.  More recent
theoretical investigations \citep{LC07} have generalized this equation
to a phenomenological relation
\begin{eqnarray}
 f(z) = \Omega_m(z)^\gamma
\end{eqnarray}
where the parameter $\gamma$, the ``gravitational growth index'' will
take on different values depending on the considered theory of
gravity, where $\gamma \approx 0.55$ for General Relativity in a
Universe dominated by a cosmological constant $\Lambda$.  In this
study we test the self-consistency of the $\Lambda$CDM cosmological
model on large scales, by measuring the growth rate from a galaxy
survey assuming a fiducial $\Lambda$CDM model and comparing these
measurements with the predicted values.

A powerful technique for determining the growth rate is to measure and
model redshift-space distortions imprinted in the 2-point galaxy
correlation function (hereafter \spcfs) of a galaxy redshift survey,
where the two dimensions ($\sigma,\pi$) correspond to galaxy pair
separations transverse and parallel to the line-of-sight.  This is
possible because the bulk flows of matter produced by large-scale
structure formation leave a pattern of correlated peculiar velocities
in the redshifts measured by galaxy surveys, introducing a
characteristic distortion into the shape of the measured \spcfs.

The models used to fit the distorted pattern present in \spcfs\ have
their origins in the works of \cite{Ka87} and \cite{H92}. The
principal ingredients which make up these models are:

\begin{itemize}

\item The underlying real-space correlation function (\cfr), which
  describes the clustering of galaxies in the absence of
  redshift-space distortions.  This function cannot be measured
  directly from real observations due to galaxy peculiar velocities,
  but can be modelled from theory or simulations.

\item A prescription for the large-scale linear distortion of the
  redshift-space correlation function due to galaxy bulk flows.

\item A function for the galaxy pairwise velocity distribution,
  $F(v)$, which represents non-linear effects in the velocity field
  resulting from small-scale interactions between galaxies and their
  surroundings.  An example of these effects is the characteristic
  spike known as the ``finger-of-god'' observed in \spcfs\ at small
  tangential separations ($\sigma < 4\, h^{-1}$ Mpc), due to the
  virialized motions of galaxies in dark matter haloes.

\item A prescription for the bias with which galaxies trace the
  underlying matter density field.

\end{itemize}

Various approaches have been used in the literature for modelling
these different components.  In the simplest version of the model a
power law is assumed for \cfr\ on small scales, an exponential
function for $F(v)$, and galaxies are assumed to trace the
matter-density field with linear bias on large scales.  This simple
model has been applied to analyses of many galaxy surveys including
recent large spectroscopic surveys such as the 2-degree Field Galaxy
Redshift Survey \citep[2dFGRS,][]{Co01} and the Sloan Digital Sky
Survey \citep[SDSS,][]{York00}.

The 2dFGRS spans an area $\approx 1500$ deg$^2$ mainly located in two
regions of sky.  \cite{Hk03} used $166{,}000$ 2dFGRS galaxies in the
redshift range $z = [0.0 , 0.2]$ to measure \spcfs\ up to separations
of $30 \, h^{-1}$ Mpc, constraining the growth rate with an accuracy
of approximately $10\%$ at an effective redshift $z = 0.15$.  The SDSS
Luminous Red Galaxy (LRG) sample contains more than $75{,}000$ objects
in the more extended redshift range $z = [0.15 , 0.47]$ spanning a
volume bigger than $1 \, h^{-3}$ Gpc$^3$.  Its correlation function
allows a measurement of $f$ at an effective redshift of $z=0.35$
\citep{CG09,Ok08,Sm12}.  Noisier measurements of the growth rate have
also been determined from higher-redshift surveys such as the VVDS
(VIMOS VLT Deep Survey) at redshift $\approx 0.8$ \citep{G08}.

Other analyses of redshift-space distortions in the galaxy
distribution, involving a wider range of cosmological parameters,
include: the studies of \cite{Pc04} who model the redshift-space
distortions of the 2dFGRS by expanding its density field in spherical
harmonics; \cite{Tm06} who measure galaxy power spectra from SDSS LRGs
at $z \sim 0.3$ and fit for a large set of cosmological parameters;
and \cite{dAn08} who use QSO catalogues from the 2QZ and 2SLAQ surveys
to compute and model \spcfs\ for $z \sim 1.5$.  These studies
generally found agreement between the measurements and the predictions
of the prevailing $\Lambda$CDM cosmological model.

Although galaxy redshift surveys currently provide the most accurate
means of measuring the cosmic growth history, other methods are also
being actively developed to extract this information.  These include
studies of the luminosity function of X-ray clusters and their
physical properties such as gas mass fraction \citep{Ra09},
measurements of bulk flows in the local neighbourhood
\citep{AL08,Wk09,Nu11}, and weak lensing distortions mapped in large,
high-quality photometric imaging surveys \citep{He07,Be09}.

In this study we analyze redshift-space distortions in the 2-point
correlation function of the WiggleZ Dark Energy Survey (hereafter
WiggleZ, \citet{Dw10}).  This is a new high-redshift survey which has
observed more than $200{,}000$ galaxy redshifts over $\sim$ 1000
deg$^2$ of sky in the redshift range $z < 1$.  The availability of the
WiggleZ data enables accurate cosmological measurements to be carried
out for the first time in the high-redshift ($z > 0.4$) Universe, the
epoch at which the Universe apparently started its accelerating
expansion.  WiggleZ galaxies are selected as star-forming (blue)
galaxies by photometric criteria \citep{Blk09,Dw10}, in contrast to
previous surveys wich targetted Luminous Red Galaxies, and therefore
also allow us to investigate if the cosmological conclusions depend on
the type of galaxy used to trace large-scale structure.

The growth rate has been measured from the WiggleZ dataset by
\citet{Chris_f} using a power-spectrum analysis.  In this study we
carry out a correlation-function analysis of a similar dataset.  There
is a strong motivation for performing both analyses.  Cosmological
measurements from large, modern galaxy surveys are limited not by
statistical errors but by systematic errors, generally arising from
modelling non-linear physics in addition to technical concerns such as
determining the data covariance matrix in a sufficiently robust and
accurate manner.  Systematics imprint themselves quite differently
into analyses in configuration space and Fourier space.  For example,
a correlation-function analysis is more effective at separating
processes which occur at different spatial scales (for example shot
noise owing to the discreteness with which galaxies trace dark matter
haloes is important at all scales in a power-spectrum measurement but
only at small scales in the corresponding correlation function).
Moreover, power-spectrum measurements are much more sensitive to how
accurately the selection function of the survey can be quantified,
which is a major issue of concern for a survey such as WiggleZ with
relatively low redshift completeness \citep{Blk09b}.  However,
modelling of quasi-linear scales becomes more robust in a
power-spectrum analysis, where linear theory can be more confidently
applied across a range of values of wavenumber $k$ (whereas structure
with a given separation in configuration space depends on a set of
underlying modes with a significant range of values of $k$).  In
general the consistency of the power-spectrum and correlation-function
analysis approaches is important to check, and would increase
confidence in the robustness of the results against systematic error.

We have investigated in detail the modelling of redshift-space
distortions in the correlation function in a companion paper
\citep{Co11a} using dark matter and halo catalogues from an N-body
simulation.  These results are important for our current study and we
summarize them here:

\begin{itemize}

\item The commonly-used power-law model for the real-space correlation
  function produces a poor fit to the clustering pattern and
  introduces considerable systematic errors in a growth rate
  measurement based on a large sample of galaxies.  We considered two
  alternative models for the real-space correlation function, which
  are able to recover the growth rate with minimal systematic error.

\item We introduced a technique which permits a non-parametric
  determination of the galaxy pairwise velocity distribution from a 2D
  correlation function.  We used this method to measure the pairwise
  velocity dispersion and check the consistency of the assumed
  functional form.

\item We explored the sensitivity of the results to the range of
  scales included in the fit, in particular considering different cuts
  for the minimum transverse scale fitted, $\sigma_{\rm min}$.

\end{itemize}

Our study of the growth rate in the WiggleZ correlation function is
structured as follows: in Section 2 we describe our data and the
measurements of \spcfs, and in Section 3 we summarize the models
fitted.  In Section 4 we determine the best-fitting model parameters
and errors, comparing these measurements to the theoretical
predictions of the $\Lambda$CDM cosmology using standard gravity.  We
also make an initial investigation of the sensitivity of our results
to the fiducial cosmological model via the Alcock-Paczynski
distortion.  In Section 5 we summarize and discuss the results.

\section{WiggleZ survey data and correlation function measurements}

\subsection{Data}
\label{sec:data}

The WiggleZ Dark Energy Survey is the latest in a series of large
spectroscopic galaxy redshift surveys mapping out large-scale
structure in the Universe.  The survey, which was carried out between
August 2006 and January 2011, obtained more than $200{,}000$ galaxy
redshifts in the range $0.1 < z < 1.0$ spread across several different
equatorial regions of sky, which together constitute a survey area of
$1000$ deg$^2$ and span a volume of approximately 1 $h^{-3}$ Gpc$^3$.
The observations were carried out at the 3.9m Anglo-Australian
Telescope (AAT) using the AAOmega spectrograph.  Individual galaxy
spectra were redshifted by WiggleZ team members.

A key characteristic of the WiggleZ Survey is the target selection
criteria \citep{Blk09,Dw10} which are based on a combination of
ultra-violet photometry from the {\em Galaxy Evolution Explorer}
(GALEX) satellite and optical imaging from the SDSS and 2nd Red
Cluster Sequence (RCS2) survey \citep{Yee05}.  WiggleZ galaxies are
star-forming (emission-line) galaxies, which are expected to avoid the
densest regions of galaxy clusters where star formation is suppressed
by mechanisms such as ram-pressure stripping of the gas content of the
galaxies.  Hence this WiggleZ dataset should be less susceptible to
the non-linear clustering effects found on small scales (such as
``fingers-of-god'') compared to surveys targetting more luminous,
highly-biased galaxies.

In this study we analyze a total of $162{,}454$ WiggleZ galaxy
redshifts, which are selected from the [1,3,9,11,15,22]-hr regions
after minor cuts for redshift and survey contiguity.  We sub-divide
the sample into four redshift slices spanning redshifts of [0.1-0.3],
[0.3-0.5], [0.5-0.7] and [0.7-0.9].  Table \ref{tab:wigg} shows the
details of the catalogues we used.  We calculate an effective redshift
$z_{\rm eff} = [0.21, 0.39, 0.61, 0.76]$ for the measurement in each
redshift slice by averaging the mean redshifts of the galaxy pairs
included in our analysis at the separations of interest.

%\begin{tabularx}{4in}{|c||c|c|c|c|c|c|c||c|}
\begin{table*}
\begin{tabular*}{0.80\textwidth}{ @{\extracolsep{\fill}} |c||c|c|c|c|c|c|c||c|}
\hline
$z_{\rm slice}$/Field & $z_{\rm eff}$ & 01 hr & 03 hr & 09 hr  &  11 hr & 15 hr  & 22 hr& \\ 
\hline
\hline
Z1 [0.1-0.3] & 0.21 & 3058 &2973 &2671  &4025  &3578  &5782& 22087\\ \hline
Z2 [0.3-0.5] & 0.39 & 2983 &3522 &7438  &9581  &10942 &7728& 42194\\ \hline
Z3 [0.5-0.7] & 0.61 & 5702 &5982 &11294 &13964 &17928 &8924& 63794\\ \hline
Z4 [0.7-0.9] & 0.76 & 3518 &4315 &5189  &7077  &8943  &5337& 34379\\ \hline \hline
Total & &  15261&16792&26592 &34647 &41391 &27771&162454\\ \hline
\end{tabular*} 
\caption{This table summarizes the number of galaxies used in our
  analysis, and the effective redshift of our measurement, by region
  and redshift slice.  The effective redshift was measured by
  averaging the mean redshift of each galaxy pair [$z_m=(z_1+z_2)/2$]
  for all the galaxy pairs counted in our 2-point correlation function
  measurements. For the scales of interest ($0-50\, h^{-1}$ Mpc) we
  find no significant dependence of $z_{\rm eff}$ on the pair
  separation.}
\label{tab:wigg}
\end{table*}

\subsection{Random catalogues}

The galaxy correlation function is measured by comparing the
pair-count of the dataset in separate bins to that of a random,
unclustered set of galaxies possessing the same selection function as
the data.  We generated random WiggleZ datasets using the method
discussed by \citet{Blk09}.  This process models several effects
including the variation of the GALEX target density with dust and
exposure time, the incompleteness in the current redshift catalogue,
the variation of that incompleteness imprinted across each 2-degree
field by constraints on the positioning of fibres and throughput
variations with fibre position, and the dependence of the galaxy
redshift distribution on optical magnitude.  Random catalogues, each
containing the same number of galaxies as the data, are generated by a
Monte Carlo process.

\subsection{Correlation function measurement}

In order to measure \spcfs\ we first converted the redshifts and
celestial coordinates of the galaxies into co-moving $(x,y,z)$ spatial
co-ordinates assuming a fiducial flat $\Lambda$CDM cosmological model
with $\Omega_m = 0.27$ (and $H_0 = 100 \, h$ km s$^{-1}$ Mpc$^{-1}$).
We investigate the sensitivity of our results to the fiducial model in
Section \ref{sec:ap}.  We then computed the number of galaxy pairs in
a series of $(\sigma,\pi)$ bins, where $\pi$ and $\sigma$ are the pair
separations parallel and transverse to the line-of-sight respectively,
where we re-define the line-of-sight direction for every pair of
galaxies using the bisector of their subtended angle on the celestial
sphere, in the same manner as \cite{H93}.  We repeated this process
using 100 random galaxy catalogues, and finally employed the Landy \&
Szalay estimator for \spcfs\ \citep{LS93}:
\begin{eqnarray}
 \xi(\sigma,\pi)=\frac{DD -2DR + RR}{RR} 
\end{eqnarray}
where $DD$ is the data-data pair count, $DR$ is the data-random pair
count, and $RR$ is the random-random pair count.  We computed these
quantities in square $(\sigma,\pi)$ bins with side $3\, h^{-1}$ Mpc
over the ranges $0 < \sigma < 30 \, h^{-1}$ Mpc, $0 < \pi < 30 \,
h^{-1}$ Mpc.  We do not use any additional weights for each data or
random object, noting that in a sample with relatively low number
density such as WiggleZ, weighting results in a negligible improvement
to the error in the correlation function over the range of scales of
interest.  The fibre-collision corrections are not important for the
scales analyzed here, and are ameliorated by repeat observations of
each patch of sky to build up the target density.  We measured
\spcfs\ in each separate WiggleZ region and redshift slice, and
optimally combined the measurements for common redshift slices using
inverse-variance weighting (see below).  Figure \ref{fig:fig1}
illustrates the combined correlation function data for each of the
four redshift slices.

\begin{figure*}
\centering
\begin{tabular}{cc}
\includegraphics[angle=-90,width=0.5\linewidth]{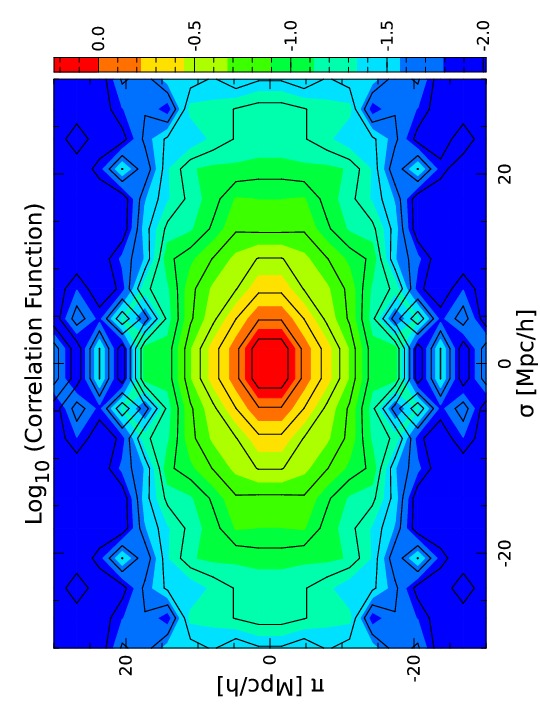} &
\includegraphics[angle=-90,width=0.5\linewidth]{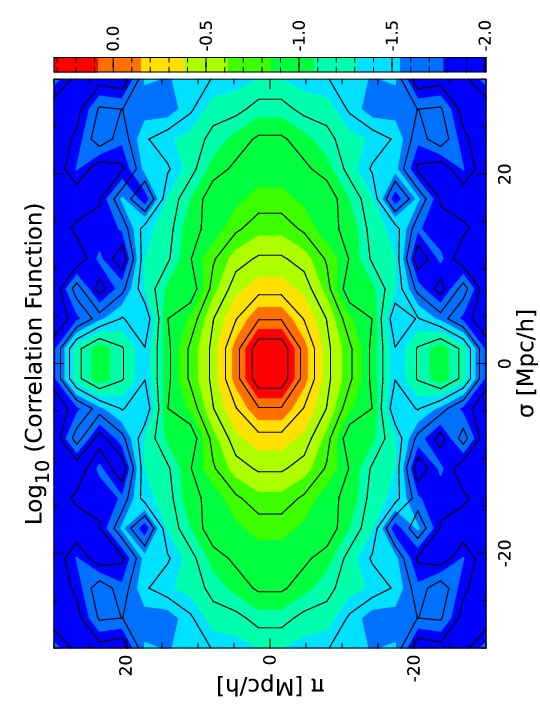} \\
\includegraphics[angle=-90,width=0.5\linewidth]{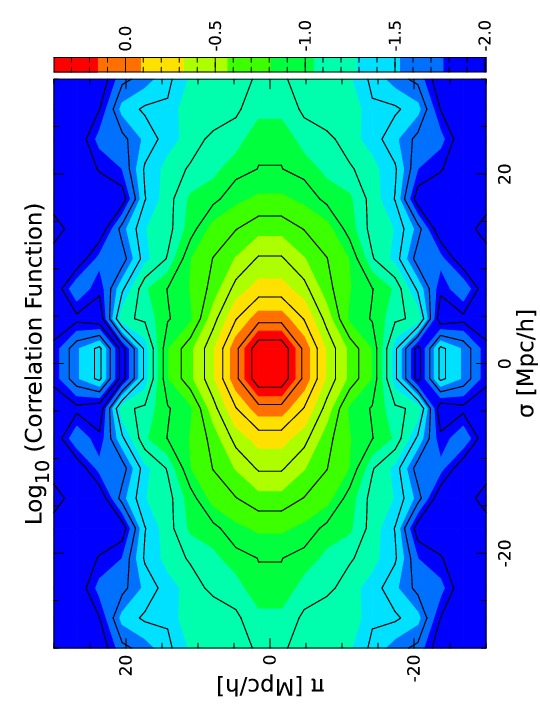} &
\includegraphics[angle=-90,width=0.5\linewidth]{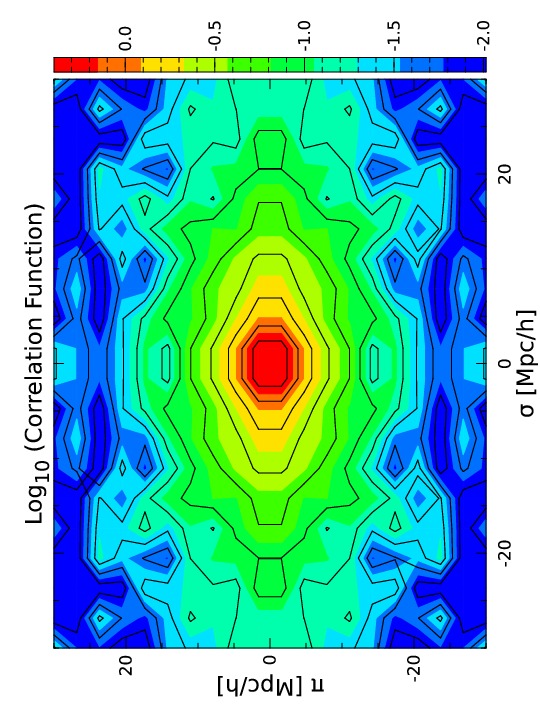}
\end{tabular}
\caption{Measurements of the redshift-space correlation function for
  redshift slices $0.1-0.3$ (top left), $0.3-0.5$ (top right),
  $0.5-0.7$ (bottom left) and $0.7-0.9$ (bottom right), obtained by
  combining results in the different WiggleZ Survey regions with
  inverse-variance weighting.  Only the top-right quadrant of data for
  each redshift is independent; the other three quadrants are mirrors
  of this first quadrant. Noticeable is the lack of ``fingers-of-god''
  prominent in similar measurements for LRGs, owing to the avoidance
  by WiggleZ galaxies of high-density regions.}
\label{fig:fig1}
\end{figure*}

\subsection{Covariance Matrix}

We used a jack-knife procedure to measure the correlation function
variances and covariances between bins.  This technique is implemented
by sub-dividing the dataset into $N_{JK}$ equal volumes, then
repeating the \spcfs\ computation $N_{JK}$ times excluding each of the
subsets in turn.  Using this procedure we obtain $N_{JK}$ different,
correlated measurements of \spcfs.  In order to compute the full
covariance matrix between the bins we use the expression
\begin{eqnarray}
  C(\xi_i, \xi_j) =  (N_{JK}-1) (\left<\xi_i\xi_j\right>-\left<\xi_i\right>\left<\xi_j\right>)
\end{eqnarray}
In our analysis we used $N_{JK} = 125$ for each individual survey
region, choosing equal-volume sub-regions tiled in three dimensions,
corresponding to a total $N_{JK} = 750$ for the combination of the 6
regions.  We checked that our results were not sensitive to these
choices.  Excluding the diagonal elements, the correlation
coefficients from the combined covariance matrices have consistently
low average values 0.10 $\pm$ 0.10; 0.09 $\pm$ 0.08; 0.07 $\pm$ 0.08
and 0.03 $\pm$ 0.07 for the four redshift slices in ascending redshift
order.
 
The combination of \spcfs\ for the 6 sky regions for every redshift
slice is produced using inverse-variance weighting:
\begin{eqnarray}
  \xi_i =  \left(\sum_{k=1}^6  \frac{\xi_{i,k}}{\sigma_{i,k}^2}\right)  \left(\sum_{k=1}^6  \frac{1}{\sigma_{i,k}^2}\right)^{-1} 
\end{eqnarray}
for the $i^{th}$ bin, and the covariance matrices are combined as
\begin{eqnarray}
  C(\xi_i,\xi_j) =  \left(\sum_{k=1}^6\frac{C(\xi_i,\xi_j)_k}{\sigma_{i,k}^2 \sigma_{j,k}^2}\right)  \left(\sum_{k=1}^6 \frac{1}{\sigma_{i,k}^2}  {\sum_{k=1}^6 \frac{1}{\sigma_{j,k}^2} }\right)^{-1} 
\end{eqnarray}
for the $i^{th}$ and the $j^{th}$ bin of \spcfs.  We determined the
bin size ($3\, h^{-1}$ Mpc) of our measurement as a compromise between
mapping the $(\sigma,\pi)$ plane with high resolution, and the need to
obtain an invertible covariance matrix.  A smaller bin size would
require a larger number of jack-knife sub-volumes to ensure a
non-singular covariance matrix, resulting in smaller jack-knife
regions which would be less independent on the scales of interest.

\section{Models for the 2D correlation function}
\label{sec:migglez}

The effect of linear redshift-space distortions on the power spectrum
in Fourier space was described by Kaiser (1987):
\begin{eqnarray}
 P_s(k,\mu) = (1 + \beta\mu^2)^2 P_r(k)
\end{eqnarray}
where $P_s$ and $P_r$ are the redshift-space and real-space galaxy
power spectra at wavenumbers $k$, $\mu$ is the cosine of the angle of
the Fourier mode to the line-of-sight, and $\beta = f/b$ quantifies
the amplitude of redshift-space distortions in terms of the growth
rate $f$ and linear bias factor $b$.  Hamilton (1992) provided the
equivalent expression in configuration-space:
\begin{eqnarray}
 \xi_s'(\sigma,\pi)=\xi_0(r)P_0(\mu)+\xi_2(r)P_2(\mu)+\xi_4(r)P_4(\mu)
\end{eqnarray}
where $P_\ell$ are the Legendre polynomials and the correlation
function multipoles $\xi_\ell(r)$ are given by the general
expressions:
\begin{eqnarray}
\label{eq:eq-gen}
\xi_0(r)=\left(1 +\frac{2\beta}{3}+\frac{\beta^2}{5}\right) \xi_r(r) \\
\xi_2(r)=\left(\frac{4\beta}{3}+\frac{4\beta^2}{7}\right)\left[\xi_r(r)-\dot{\xi}_r(r) \right] \\
\xi_4(r)=\left(\frac{8\beta^2}{35}\right)  \left[\xi_r(r)+\frac{5}{2}\dot{\xi}_r(r)-\frac{7}{2}\ddot{\xi}_r(r) \right] 
\end{eqnarray}
in terms of the integrals
\begin{eqnarray}
\dot{\xi}_r(r)=\frac{3}{r^3}\int_0^r{\xi_r(s) s^2 ds} \\
\ddot{\xi}_r(r)=\frac{5}{r^5}\int_0^r{\xi_r(s) s^4 ds} . 
\end{eqnarray}
In order to complete the model, the 2D correlation function including
linear redshift-space distortions, $\xi_s'(\sigma,\pi)$, is then convolved with a function
$F(v)$ representing a dispersion of pairwise velocities on small
scales:
\begin{eqnarray}
 \xi_s(\sigma,\pi)=\int^{\infty}_{-\infty}\xi_s'\left(\sigma,\pi-\frac{v}{H(z)a(z)}\right) F(v) \, dv
\end{eqnarray}
normalized such that $\int^{\infty}_{-\infty} F(v) \, dv = 1$.

The most common approach in previous analyses of redshift-space
distortions in the galaxy correlation function has been to use a
power-law choice for \cfr\ and an exponential 1D velocity distribution
for $F(v)$ \citep{LS98,Hk03,CG09}.  In a companion paper \citep{Co11a}
we presented a detailed analysis of the systematic errors resulting
from these assumptions, considering three different models for the
underlying real-space correlation function \cfr:

\begin{itemize}

\item {\it Power-law model:}
  $\xi_r(r)=\left(\frac{r}{r_0}\right)^{-\gamma }$.

\item {\it CAMB model:} we use a real-space correlation function
  calculated numerically for a given set of cosmological parameters
  (motivated by analyses of the Cosmic Microwave Background radiation)
  using the CAMB software \citep{LC99} with the WMAP-7 best-fit
  cosmological parameters.

\item {\it Quadratic Correlation Function (QCF) model:} we generalize
  the power-law function to an empirical real-space correlation with
  one further degree of freedom: $\xi_r(r) =
  \left(\frac{r}{r_0}\right)^{-\gamma + q
    log_{10}\left(\frac{r}{r_0}\right)}$.  This form has more
  flexibility to describe the real-space correlation function than the
  other models. For $q=0$ the QCF recovers the power-law, while for
  $q\sim-0.5$ the QCF is similar to a CAMB correlation function.

\end{itemize}

We note two advantages of the QCF model.  Firstly it does not assume a
particular set of cosmological parameters, so represents a more
flexible, model-independent approach that is able to reproduce the
shape of the CAMB correlation function for a range of cosmological
parameters.  Secondly, the flexibility of the QCF model allows it to
accommodate effects such as scale-dependent bias, which is expected to
be important for massive haloes hosting Luminous Red Galaxies.

In \citet{Co11a} we also considered a series of options for the shape
of the pairwise velocity distribution function $F(v)$, introducing a
parameter $x$ which interpolates between exponential and Gaussian
shapes for this function:
\begin{eqnarray}
F(v) = x F(v)_e + (1-x) F(v)_g
\end{eqnarray}
The motivation for adding this parameter was to allow the data to
determine which of the exponential or Gaussian distributions provided
the best fit.  We found that, when fitting to halo catalogues from
dark matter simulations, the small-scale non-linear anisotropy in the
clustering pattern favoured the exponential form, whilst if we
excluded the small-scale part of the data by restricting the fitted
range to $\sigma > \sigma_{\rm min}$ the choice of exponential or
Gaussian was rendered unimportant.  In our study of N-body simulations
we were able to measure the shape of the pairwise velocity
distribution directly, and the results were self-consistent with the
best-fitting model for $F(v)$ when the CAMB or QCF models were used
for the real-space correlation function. When the power-law model was
used for $\xi_r(r)$ instead, we found systematically higher values for
the pairwise velocity dispersion. This result provided more evidence
that using an accurate model for the real-space correlation function
yielded constraints for all the fitted variables with lower systematic
error.

We reformulated the problem of obtaining the pairwise velocity
distribution by the process of fitting a non-parametric stepwise
function instead of a exponential or Gaussian profile. We introduced a
fast technique to do this \citep[][appendix A]{Co11a}.  The results
for the pairwise velocity dispersion measured as a stepwise function
from the simulation were consistent with both the direct measurements,
and with the fits of parameterized models.

Finally, we concluded from our experiments with simulations that our
models were still not complex enough to model the small-scale
clustering pattern of dark matter halo catalogues, and we were
required to introduce a cut $\sigma > \sigma_{\rm min} \sim 6 \, h^{-1}$
Mpc in the fitted range to restrict systematic errors in the derived
growth rate to a maximum of $\Delta f \approx 0.05$.

\section{Results}

\subsection{Fitting procedure}

We fitted the models described in Section \ref{sec:migglez} to the
WiggleZ dataset by minimizing the $\chi^2$ statistic calculated as
\begin{eqnarray}
\chi^2 = \sum_{ij} (M_i - D_i)^T C^{-1}_{ij}(M_j-D_j)
\end{eqnarray}
where $M_i$ is the model vector for $i$ data bins, $D_i$ is the data
vector, and $C^{-1}_{ij}$ is the element of the inverse covariance
matrix which relates bins $i$ and $j$.  Our default redshift-space QCF
model contained 6 parameters $[r_0, \gamma, q, \beta, \sigma_v, x]$
and we considered variations as described in the text.  We used a
Monte Carlo Markov Chain (MCMC) code to perform these fits,
determining the best-fitting values and joint likelihood distributions
of each parameter.  We found that chains of $100{,}000$ iterations
gave robust results for all models.  We double-checked our results
using downhill simplex algorithms.  Following \citet{Co11a}, we
investigated the dependence of our results on the minimum transverse
scale $\sigma$ included in our fits.  The correlation function at low
$\sigma$ may contain the signature of strong non-linearities not fully
described by our models, potentially introducing a systematic error in
the growth measurement.

\subsection{Measurements of the redshift-space distortion parameter $\beta$}

\begin{figure}
\centering
\includegraphics[width=95mm,angle=-90]{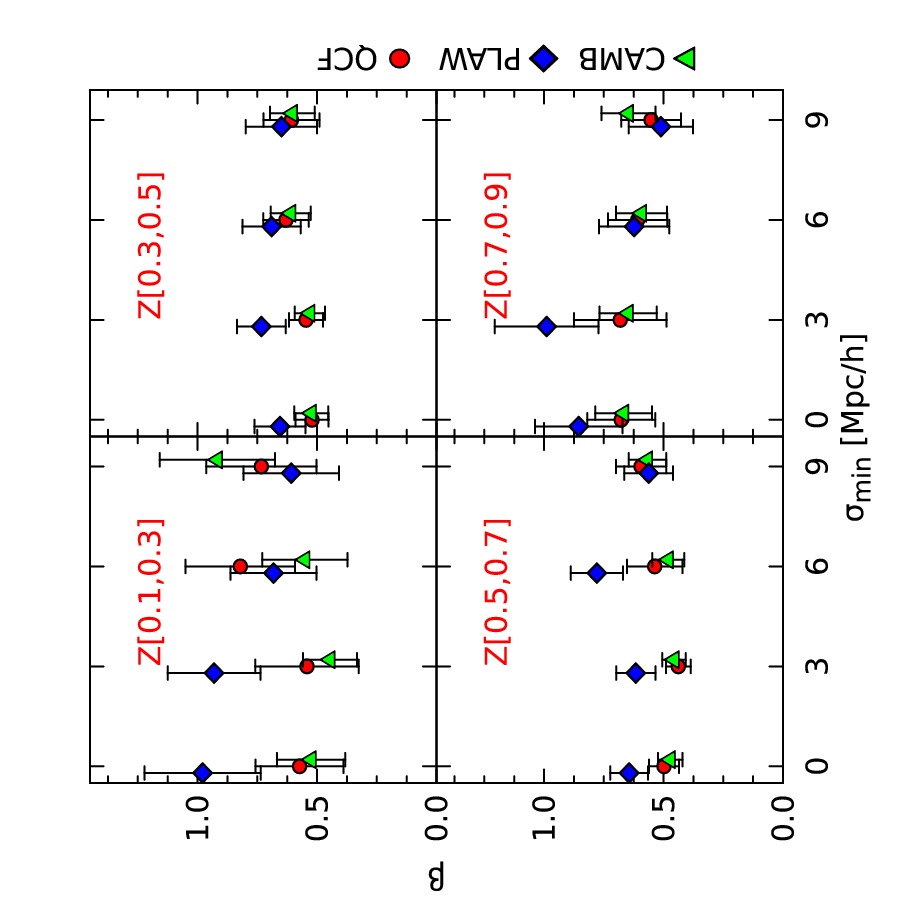}
\newline
\caption{Measurements of the redshift-space distortion parameter
  $\beta$ for the four WiggleZ redshift slices, fitting to the
  combined correlation function of the survey regions.  We show
  results for the three real-space correlation function models
  discussed in the text: QCF, CAMB and power-law, as a function of the
  minimum transverse scale $\sigma_{\rm min}$ included in the fit.
  Note the convergence of the three correlation-function models after
  small-scale bins are excluded from the fits.}
\label{fig:beta}
\end{figure}

In Figure \ref{fig:beta} we illustrate the measurements of $\beta$ in
each redshift slice, combining results from the different WiggleZ
regions as a function of the minimum transverse separation
$\sigma_{\rm min}$.  We show results for each of the three real-space
correlation function models introduced above.  We found signatures of
systematic errors at small scales with the same pattern as found in
our investigation of mock halo catalogues from N-body simulations,
especially when the power-law model was assumed.

\subsection{Measurements of the galaxy bias factor $b$}

We determined the linear galaxy bias factor $b$, which should be
combined with the measurement of $\beta$ to produce the growth rate $f
= \beta\, b$, assuming the CAMB correlation function model
describes well the dark matter correlation function $\xi_m$. Then, the
bias is determined simultaneously with the other parameters when
fitting the CAMB model to the galaxy correlation function $\xi_g$
using:
\begin{eqnarray}
\xi_g = b^2 \, \xi_m . 
\label{eqn:bias}
\end{eqnarray}
Figure \ref{fig:bias} illustrates the best-fitting bias in each
redshift slice as a function of $\sigma_{\rm min}$.

\begin{figure}
\centering
\includegraphics[width=90mm,angle=-90]{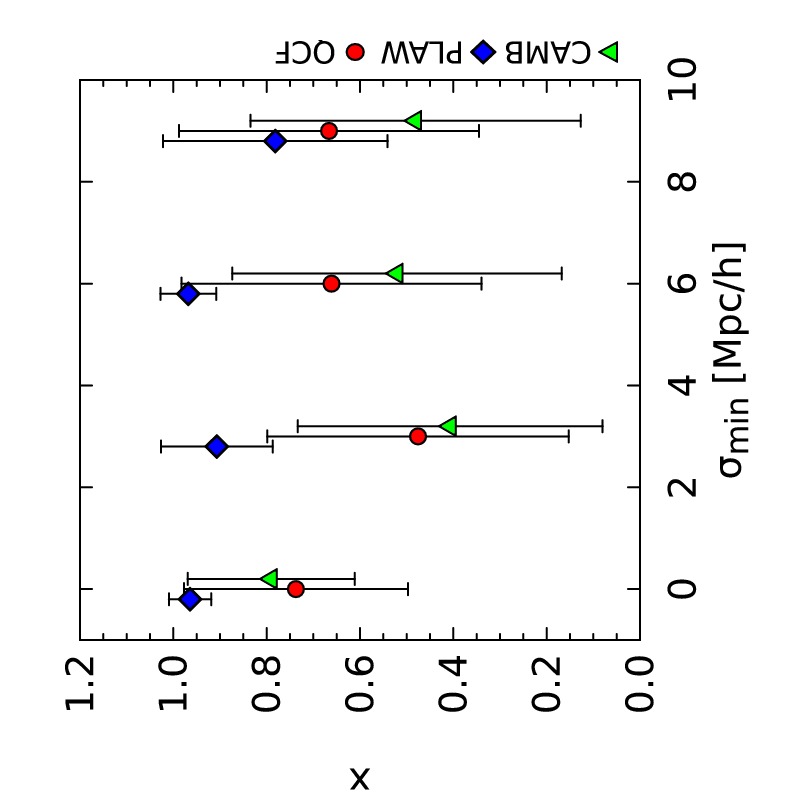}
\newline
\caption{Example of the behaviour of the parameter $x$
as a function of the minimum transverse scale
$\sigma_{\rm min}$ included in the fit, in this case
for the third redshift slice. Values of $x$ close to
1 implies that $F(v)$ has an exponential shape rather 
than a Gaussian shape. It becomes less well determined when 
excluding the small-scale part of the data.}
\label{fig:x}
\end{figure}

\begin{figure}
\centering
\includegraphics[width=95mm,angle=-90]{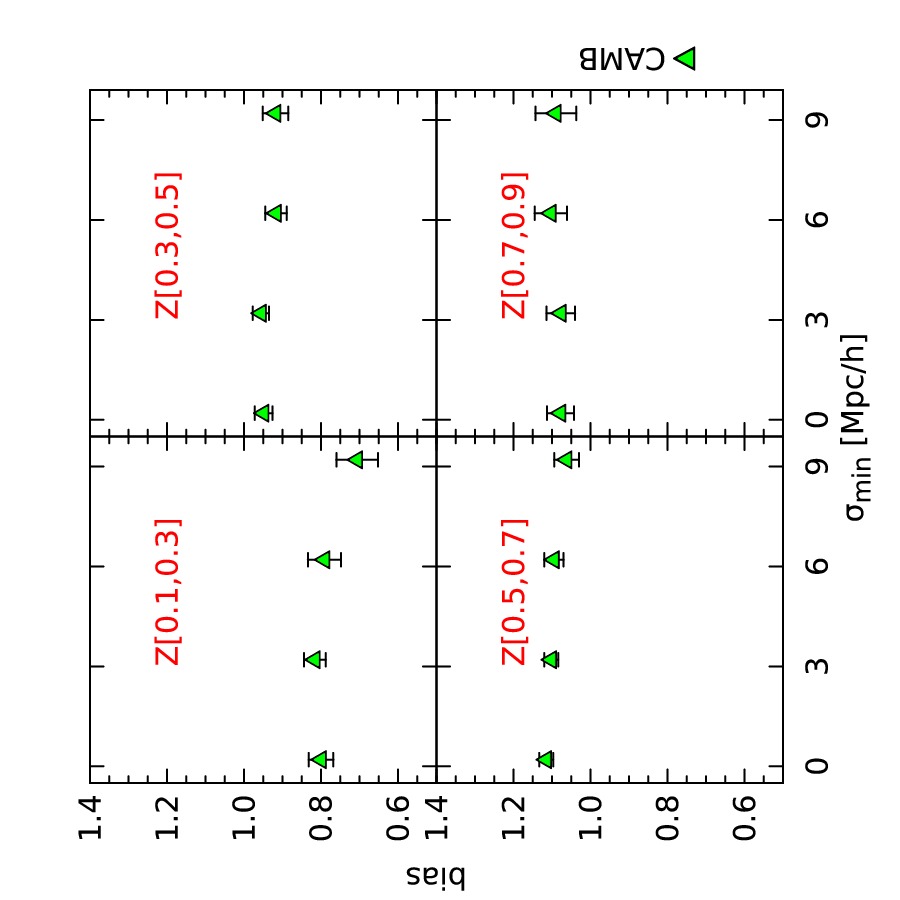}
\newline
\caption{Measurements of the galaxy bias factor in the four WiggleZ
  redshift slices as a function of the minimum transverse scale
  $\sigma_{\rm min}$ included in the fit.  The bias is determined by
  fitting a CAMB model to the 2D correlation function.  We note that
  the bias measurements presented in this Figure have not been
  corrected for the effect of redshift blunders (see text).}
\label{fig:bias}
\end{figure}

The results for $x$ and $\sigma_v$ show similar behaviour as found in
our simulation study: for small $\sigma_{\rm min}$ the value of $x$
favoured an exponential shape for $F(v)$ and $\sigma_v$ was tightly
constrained, and for larger $\sigma_{\rm min}$ both parameters were
not well-determined. See Figure \ref{fig:x} for an example.

\begin{figure}
\centering
\includegraphics[width=95mm,angle=-90]{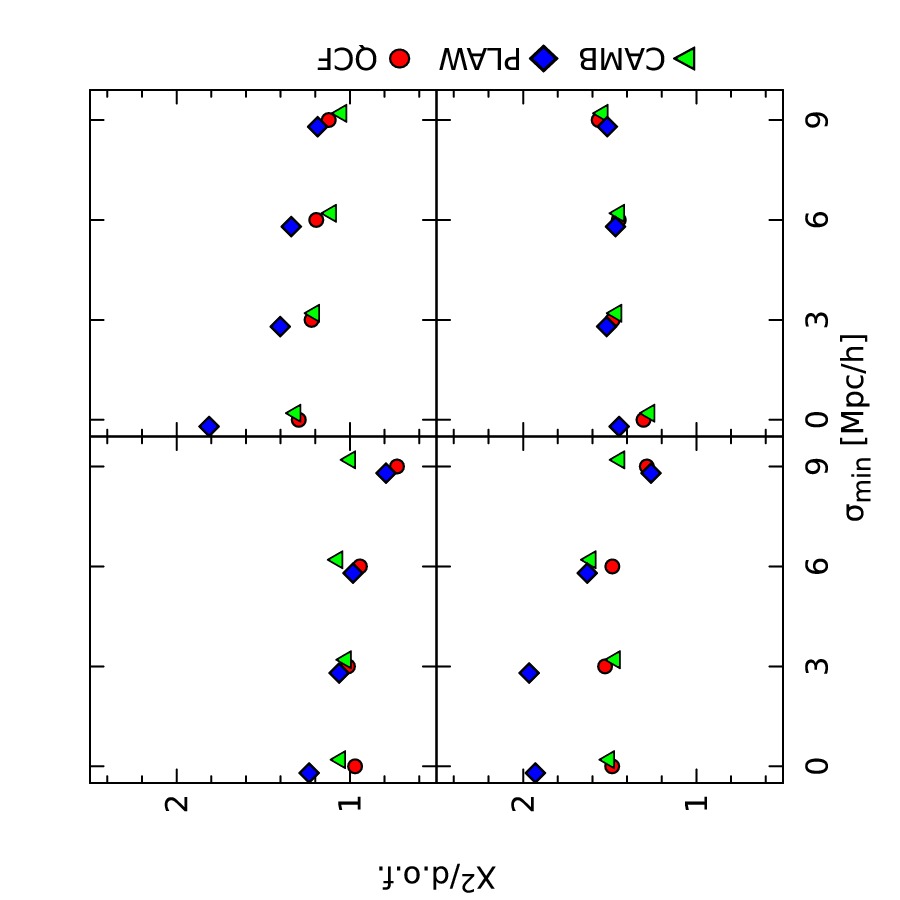}
\newline
\caption{Reduced $\chi^2$ values corresponding to the best-fitting
  parameters of each of the correlation function models in the four
  WiggleZ redshift slices, as a function of the minimum transverse
  scale.}
\label{fig:xi}
\end{figure}

The reduced $\chi^2$ values of the best-fitting models are displayed
in Figure \ref{fig:xi}.  We note that the QCF and CAMB models both
produce good fits to the data, with the QCF model carrying the
additional advantage of not assuming a small-scale shape for the
correlation function.

%\newpage
The fitted galaxy bias factors must be corrected for the (relatively
small) effects of redshift blunders caused by the mis-identification
of emission lines in the WiggleZ survey scattering the true position
of galaxies.  This effect has been carefully quantified by
\citet[][Section 3.4]{Blk09} and causes a loss of signal in the
correlation function which may be recovered by multiplying by a
redshift-dependent correction factor:
\begin{eqnarray}
\xi(s)_{corr} = \xi(s) \, [1 - r(z)]^{-2}
\end{eqnarray}
where $r$ is the fraction of redshift blunders at redshift $z$.  The
relevant correction factors $(1-r)^{-2}$ for the different redshift
slices can be found in Table \ref{res}, and are equivalent to an
upward correction in the measured galaxy bias factors.

\subsection{Stepwise velocity distribution}

\begin{figure}
\centering
\includegraphics[width=95mm]{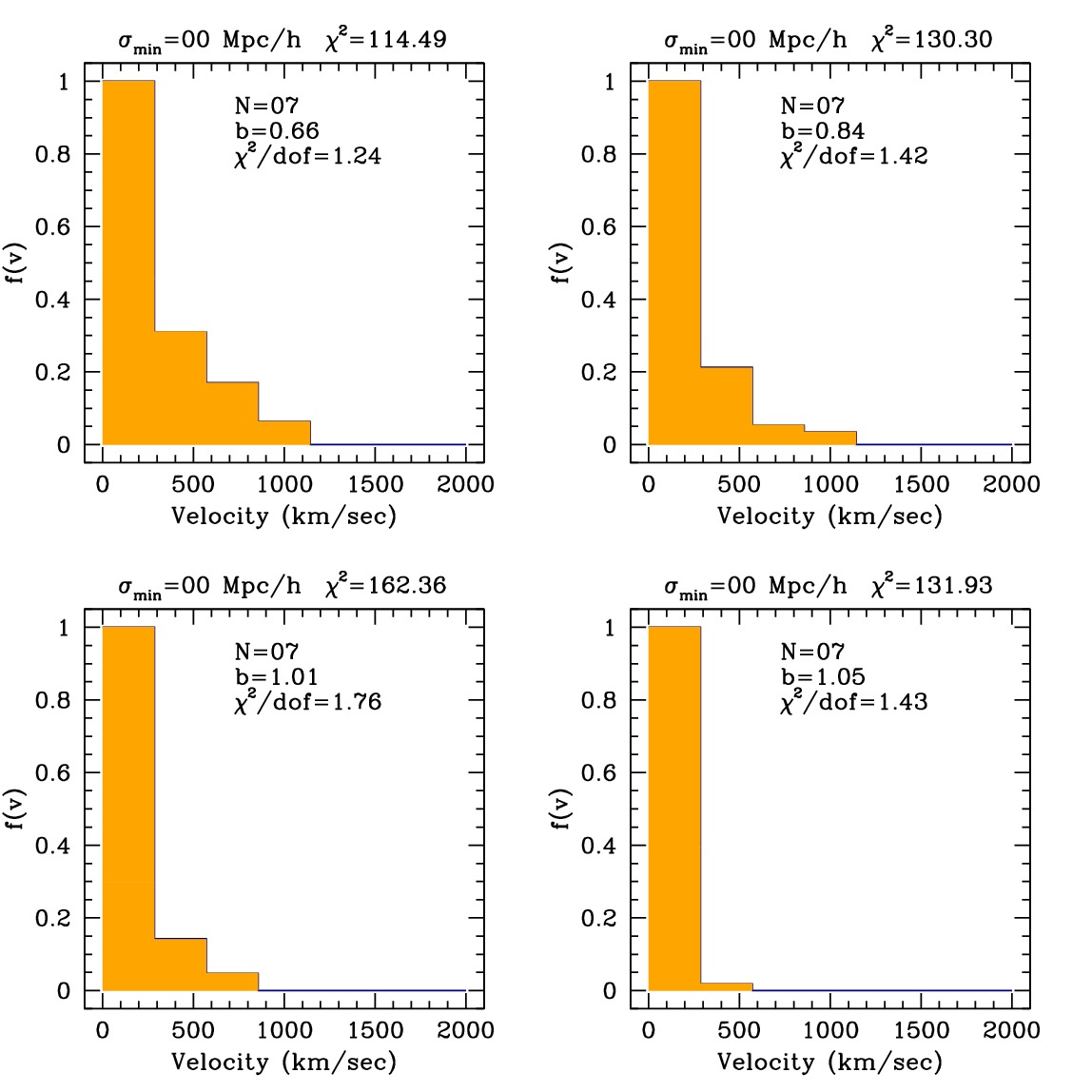}
\newline
\caption{The best-fitting stepwise values of the pairwise velocity
  distribution of WiggleZ galaxies.  We find that the amplitude of the
  pairwise velocities systematically decreases with increasing
  redshift, which is expected in a model where non-linear structure
  grows with time.}
\label{fig:sw}
\end{figure}

Our stepwise fitting technique allows us to measure the pairwise
velocity distribution in a non-parametric fashion, instead of fixing
its shape to some functional form.  The 2D correlation function of
WiggleZ galaxies does not contain strong ``fingers-of-god'' features
such as those encountered in redder galaxy samples such as SDSS LRGs.
This is expected, as WiggleZ galaxies are blue star-forming galaxies,
which inhabit less dense environments than their red counterparts and
are thereby less affected by non-linear clustering.  Figure
\ref{fig:sw} shows the non-parametric stepwise velocity distribution
obtained in every redshift slice of our sample.  We used seven bins in
velocity of width $\sim$ 300 km s$^{-1}$ up to a maximum of 2000 km
s$^{-1}$.  We find a slow growth in the width of the pairwise velocity
distribution as redshift decreases, as expected due to the growth of
non-linear structure with time. Where the data permits us to
discriminate, the measured stepwise distribution is consistent with an
exponential function for $F(v)$.

\subsection{Measurements of the growth rate $f$}

\begin{figure}
\centering
\includegraphics[width=95mm,angle=-90]{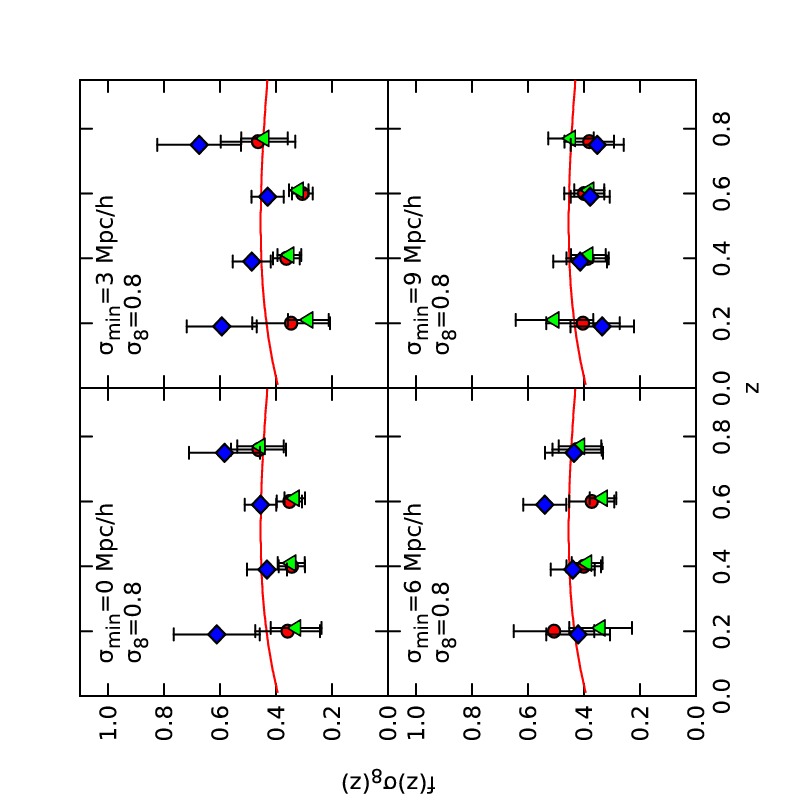}
\newline
\caption{The red circles, green triangles and blue diamonds represent
  the fitted values of the growth rate in the four WiggleZ redshift
  slices using a QCF, CAMB and power-law model respectively for
  different data cuts $\sigma > \sigma_{\rm min}$.  This plot shows
  the convergence of the three models when more of the small-scale
  part of the data is excluded.  We note that our simulation study
  \cite{Co11a} shows that the growth rate is systematically
  under-estimated if $\sigma_{\rm min}$ is too low, with $\sigma_{\rm
    min} \sim 6 \, h^{-1}$ Mpc required to obtain systematic-free
  results with the QCF model.  In all panels we overplot the
  prediction of a fiducial $\Lambda$CDM model with $\Omega_m = 0.27$
  and $\sigma_8 = 0.8$.}
\label{fig:gr}
\end{figure}

The final growth rate measurements are obtained by combining $\beta$
and the bias: $f(z) = \beta \, b$ .  However, as the bias was measured
assuming a CAMB matter correlation function with a fixed normalization
$\sigma_8$, the measured value of $b$ is degenerate with $\sigma_8$.
We can convert our measurements to be independent of this effect by
quoting the value of $f(z) \sigma_8(z)$ \citep{So09}.  In Figure
\ref{fig:gr} we display these results in the four redshift slices for
different choices of the fitting range $\sigma_{\rm min}$.  Motivated
by the simulation results presented by \cite{Co11a}, we take as our
preferred measurement the QCF model with $\sigma_{\rm min} = 6 \,
h^{-1}$ Mpc.

\begin{figure}
\centering
\includegraphics[width=70mm,angle=-90]{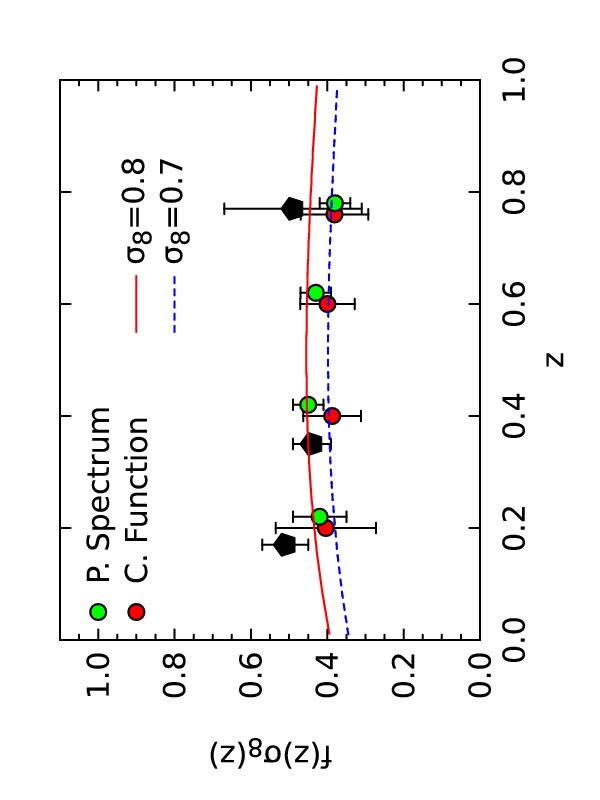}
\newline
\caption{Growth rate results from power-spectrum and
  correlation-function analyses (green and red circles) of WiggleZ
  data. In black, we overplot results from the 2dFGRS, SDSS and VVDS
  surveys (in increasing redshift order).  Both WiggleZ sets of
  results are shifted in redshift for clarity.  We show predictions
  for $\Lambda$CDM cosmological models assuming $\Omega_m = 0.27$ and
  $\sigma_8 = 0.7$ (dashed line) and $0.8$ (solid line).}
\label{fig:GR}
\end{figure}

In Figure \ref{fig:GR} we compare these preferred measurements to the
results from the analysis of the 2D WiggleZ survey power spectra
\citep{Chris_f}, together with some previous galaxy surveys, and
overplot the prediction of a $\Lambda$CDM model with matter density
$\Omega_m = 0.27$ and two different values of the low-redshift
normalization $\sigma_8 = 0.7$ and $0.8$.  The separate measurements
using different datasets and techniques are consistent within the
statistical errors.  The errors in the growth rate measurements using
the 2D WiggleZ correlation function are around twice the size of those
determined by fits to the power spectra \citep{Chris_f}.  We attribute
this to a combination of the greater number of parameters that must be
varied to obtain a good fit to the correlation function given the
greater relative importance of the non-linear effects that must be
modelled, and the more limited range of scales $(\sigma, \pi) < 30 \,
h^{-1}$ Mpc studied.  Nonetheless, the statistical agreement of these
analyses is evidence that the different measurements of the growth
rate are not dominated by systematic errors.

\subsection{Effects of the Alcock-Paczynski distortion}
\label{sec:ap}

Finally we tested the sensitivity of our results to the fiducial
cosmological model adopted in our analysis by varying the value of
the matter density $\Omega_m$.  The sensitivity of our measurements to
the fiducial model is a result of the Alcock-Paczynski effect
\citep{AP79}, a geometrical distortion of the measured clustering
pattern which arises if the trial cosmology differs from the true
cosmology.  The effect is partially degenerate with redshift-space
distortions \citep{Bll96,Ma96,Ma00,Seo03,Smps10} resulting in
systematic variations in the best-fitting growth rate as the fiducial
cosmology is altered.  Alternatively, assuming that the cosmological
model is known from a $\Lambda$CDM standard gravity model with
parameters tuned to match the Cosmic Microwave Background
anisotropies, the Alcock-Paczynski effect provides a further
cross-check of this model.

\begin{figure}
\centering
\includegraphics[width=90mm,angle=-90]{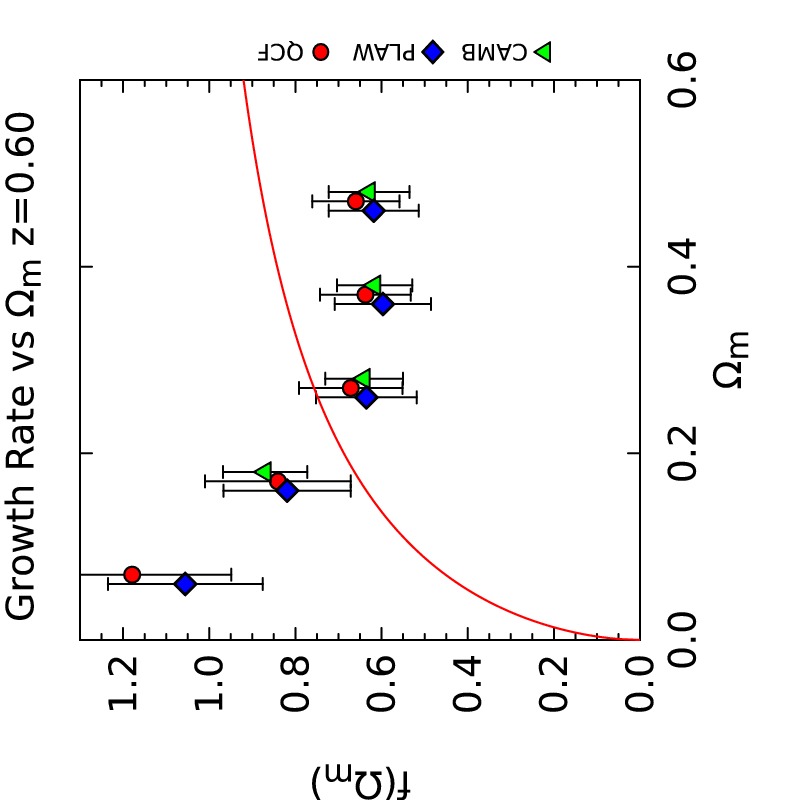}
\newline
\caption{A self-consistency test of the $\Lambda$CDM cosmology
  assuming standard gravity, in which the WiggleZ growth rate
  measurements in the $z=0.6$ redshift slice are repeated for
  different fiducial choices of $\Omega_m$.  The results are sensitive
  to the Alcock-Paczynski effect, and produce a measured growth rate
  inconsistent with the prediction of standard gravity for both low
  and high values of $\Omega_m$.}
\label{fig:ap}
\end{figure}

We implemented this analysis by repeating our fitting procedure for
different values of the fiducial matter density $\Omega_m = (0.07,
0.17, 0.37, 0.47)$ for the central redshift slice $z = 0.6$.  The
results are shown in Figure \ref{fig:ap}, demonstrating that the
$\Lambda$CDM WMAP-cosmological model is self-consistent in the range
$0.2 < \Omega_m < 0.3$ and produces inconsistent results (in standard
gravity) for other values of $\Omega_m$.  We note however that our
treatment here is approximate; in our power spectrum model we have
fixed the values of the other cosmological parameters at the WMAP7
best-fitting values.  A full treatment of these model fits is beyond
the scope of the current study and is provided by Parkinson et
al.\ (in prep.).

\section{Conclusions}

We have analysed redshift-space distortions in the 2-point galaxy
correlation function of the WiggleZ Dark Energy Survey built from over
$160{,}000$ galaxies, dividing the dataset into four redshift slices.
We fitted a series of different models to the observed \spcfs\ (2D
2-point correlation function) using a Monte Carlo Markov Chain
process.  We summarize our results as follows:

\begin{itemize}

\item We determine the cosmic growth rate across the redshift range
  $0.1 < z < 0.9$ with an accuracy of around $20\%$ in each redshift
  slice of width $\Delta z = 0.2$, and our measurements are consistent
  with the prediction of a $\Lambda$CDM model with $\Omega_m \approx
  0.27$ and $\sigma_8 \approx 0.8$.

\item Our results are also consistent with an independent
  power-spectrum analysis of a similar dataset \citep{Blk11b},
  demonstrating that systematic errors are not significant in these
  two different approaches.

\item We perform a non-parametric determination of the pairwise
  velocity distribution of WiggleZ galaxies as a stepwise function,
  demonstrating that the amplitude of pairwise velocities grows with
  decreasing redshift, as expected due to the growth of non-linear
  structure with time, and that (where it is possible to fit a model)
  it is well-described by an exponential function.

\item Our measurements agree well with the behaviour found in our
  previous experiments using N-body simulations \citep{Co11a}, and we
  obtain a convergence in the different models for the growth rate
  when excluding the small-scale non-linear part of the data ($\sigma
  < 6 \, h^{-1}$ Mpc).

\item We repeated the entire procedure (from transforming redshift
  catalogues into 3D spatial galaxy distributions to running MCMC
  correlation function model-fitting) for $\Lambda$CDM cosmologies
  with different values of $\Omega_m$. The results show that the
  $\Lambda$CDM model is only self-consistent in the range $0.2 <
  \Omega_m < 0.3$.

\end{itemize}

We conclude that the standard $\Lambda$CDM cosmological model provides
a good description of the growth rate which drives redshift-space
distortions in the clustering of WiggleZ galaxies, using cosmological
parameters which also yield a good simultaneous fit to the expansion
history measured by baryon acoustic oscillations and supernovae, and
to the fluctuations in the Cosmic Microwave Background radiation.  We
have shown that different statistical analyses of the WiggleZ dataset,
with very different potential systematic errors, produce growth rate
measurements in close mutual agreement.

\section*{Acknowledgements}

We acknowledge financial support from the Australian Research Council
through Discovery Project grants which have funded the positions of
MP, GP, TD and FM.  SMC acknowledges the support of the Australian
Research Council through a QEII Fellowship.  MJD and TD thank the
Gregg Thompson Dark Energy Travel Fund for financial support.  CC
thanks David Parkinson for sharing his expertise on MCMC techniques,
and Ana Mar\'ia Mart\'inez for her invaluable feedback and support in
the building of this paper.

GALEX (the Galaxy Evolution Explorer) is a NASA Small Explorer,
launched in April 2003.  We gratefully acknowledge NASA's support for
construction, operation and science analysis for the GALEX mission,
developed in co-operation with the Centre National d'Etudes Spatiales
of France and the Korean Ministry of Science and Technology.

Finally, the WiggleZ survey would not be possible without the
dedicated work of the staff of the Australian Astronomical Observatory
in the development and support of the AAOmega spectrograph, and the
running of the AAT.

This research was supported by CAASTRO: http://caastro.org .

\bibliography{mybib}
\bibliographystyle{mn2e}

\begin{sideways}
\label{res}
\begin{tabular}{|c|c|c|c|c|c|c|c|c|c|c|c|c|c|}
\hline
redshift&$\sigma_{\rm min}$ &\multicolumn{2}{c|}{QCF Model}&\multicolumn{2}{c|}{CAMB Model}&\multicolumn{2}{c|}{Power-Law Model} & bias &$\sigma_8(z)$&$z$-blunder&QCF &CAMB &PLAW \\
slice & $h^{-1}$ Mpc & $\beta$ $(\Delta_{\beta})$  & $\chi^2/{\rm dof}$ &$\beta$ $(\Delta_{\beta})$ & $\chi^2/{\rm dof}$ &$\beta$ $(\Delta_{\beta})$ &$\chi^2/{\rm dof}$ & $b$ $(\Delta_{b})$ & & corr. &$f\sigma_8$ & $f\sigma_8$ & $f\sigma_8$\\
\hline
\hline

\multirow{4}{*}{1}& 00 & 0.57(0.18)&0.970&0.53(0.14)&1.055&0.98(0.24)&1.235&0.80(0.03)&\multirow{4}{*}{0.728}&\multirow{4}{*}{1.067}&0.36(0.11)&0.33(0.09)&0.61(0.15) \\ 
		   & 03 & 0.54(0.22)&1.011&0.45(0.11)&1.022&0.93(0.19)&1.062&0.82(0.03)&		      &		      &0.34(0.14)&0.28(0.07)&0.59(0.12) \\ 
		   & 06 & 0.82(0.23)&0.942&0.55(0.18)&1.071&0.68(0.18)&0.982&0.79(0.04)&		      &		      &0.50(0.14)&0.34(0.11)&0.42(0.11) \\ 
		   & 09 & 0.73(0.23)&0.728&0.92(0.24)&0.997&0.61(0.20)&0.791&0.71(0.05)&		      &		      &0.40(0.13)&0.50(0.14)&0.33(0.11) \\ \hline 
\multirow{4}{*}{2}& 00 & 0.52(0.07)&1.295&0.52(0.07)&1.313&0.66(0.11)&1.813&0.95(0.02)&\multirow{4}{*}{0.662}&\multirow{4}{*}{1.050}&0.34(0.05)&0.35(0.05)&0.43(0.07) \\ 
		   & 03 & 0.55(0.07)&1.221&0.53(0.06)&1.205&0.73(0.10)&1.402&0.96(0.02)&		      &		      &0.36(0.05)&0.35(0.04)&0.49(0.07) \\ 
		   & 06 & 0.63(0.10)&1.194&0.61(0.08)&1.108&0.69(0.12)&1.338&0.92(0.03)&		      &		      &0.40(0.06)&0.39(0.05)&0.44(0.08) \\ 
		   & 09 & 0.61(0.12)&1.122&0.60(0.09)&1.048&0.65(0.15)&1.186&0.92(0.03)&		      &		      &0.39(0.08)&0.38(0.06)&0.41(0.10) \\ \hline
\multirow{4}{*}{3}& 00 & 0.50(0.06)&1.486&0.47(0.05)&1.503&0.64(0.08)&1.929&1.11(0.02)&\multirow{4}{*}{0.595}&\multirow{4}{*}{1.064}&0.35(0.04)&0.33(0.04)&0.46(0.06) \\ 
		   & 03 & 0.44(0.05)&1.527&0.46(0.05)&1.470&0.62(0.08)&1.965&1.10(0.02)&		      &		      &0.31(0.04)&0.32(0.03)&0.43(0.06) \\ 
		   & 06 & 0.54(0.12)&1.485&0.48(0.07)&1.610&0.78(0.11)&1.630&1.09(0.03)&		      &		      &0.37(0.08)&0.33(0.05)&0.54(0.08) \\ 
		   & 09 & 0.59(0.10)&1.286&0.57(0.08)&1.444&0.56(0.10)&1.262&1.06(0.03)&		      &		      &0.40(0.07)&0.38(0.05)&0.38(0.07) \\ \hline
\multirow{4}{*}{4}& 00 & 0.68(0.14)&1.305&0.67(0.12)&1.270&0.85(0.18)&1.446&1.08(0.04)&\multirow{4}{*}{0.553}&\multirow{4}{*}{1.145}&0.46(0.10)&0.46(0.08)&0.58(0.13) \\ 
		   & 03 & 0.68(0.19)&1.483&0.65(0.12)&1.461&0.99(0.22)&1.518&1.08(0.04)&		      &		      &0.46(0.13)&0.44(0.08)&0.67(0.15) \\ 
		   & 06 & 0.61(0.12)&1.448&0.59(0.11)&1.445&0.62(0.15)&1.467&1.10(0.04)&		      &		      &0.42(0.09)&0.41(0.08)&0.44(0.10) \\ 
		   & 09 & 0.55(0.12)&1.564&0.65(0.11)&1.540&0.51(0.13)&1.514&1.09(0.05)&		      &		      &0.38(0.09)&0.45(0.08)&0.35(0.09) \\ \hline \hline
\multicolumn{14}{l}{{\bf Table 2.} The best-fitting values and errors for the most important parameters of the redshift-space distortion models fitted in each WiggleZ redshift slice.} \\ \hline
%\caption{The best-fitting values and errors for the most important parameters of the redshift-space distortion models fitted in each WiggleZ redshift slice.}
\end{tabular}
\end{sideways}

\end{document}